\documentclass[aip]{revtex4-1} 
\usepackage{amsmath}
\usepackage{svg}
\usepackage[hangindent=2em, labelfont=bf, font=small,]{caption}
\usepackage{subcaption}
\usepackage{float}
\usepackage[utf8x]{inputenc}

\newcommand{\unit}[1]{\mathrm{\, #1}}

\newcommand{\mux}[1]{\mu_\mathrm{#1}}

\begin{document}

\title{Machine learning for ultra high throughput screening of organic solar cells: Solving the needle in the hay stack problem.}

\author{Markus Hu{\ss}ner}

\affiliation{Department of Engineering, Durham University, Lower Mount Joy, South Road, Durham, DH1 3LE, UK}

\author{Pacalaj A. Richard}
\affiliation{Department of Chemistry and Centre for Processable Electronics"
MSRH, 82 Wood Lane, W12 0BZ}

\author{Olaf G. Müller-Dieckert}
\affiliation{Institut für Physik, Technische Universität Chemnitz, Straße der Nationen 62, D-09111 Chemnitz, Germany}

\author{Chao Liu}
\affiliation{Department of Materials  Science and Engineering, Friedrich-Alexander-Universität Erlangen-Nürnberg, 91054 Erlangen, Germany}

\author{Zhisheng Zhou}
\affiliation{State Key Laboratory of Luminescent Materials and Devices, School of Materials Science and Engineering, South China University of Technology, Guangzhou, 510640, China}

\author{Nahdia Majeed}
\affiliation{Faculty of Engineering, The University of Nottingham, University Park, Nottingham, NG7 2RD}

\author{Steve Greedy}
\affiliation{Faculty of Engineering, The University of Nottingham, University Park, Nottingham, NG7 2RD}

\author{Ivan Ramirez}
\affiliation{Heliatek GmbH, Dresden, Treidlerstraße 3, 01139 Dresden, Germany}

\author{Ning Li}
\affiliation{State Key Laboratory of Luminescent Materials and Devices, Institute of Polymer Optoelectronic Materials and Devices, School of Materials Science and Engineering, South China University of Technology, Guangzhou, 510640 China}
\affiliation{Helmholtz Institute Erlangen-Nürnberg for Renewable Energy (HI ERN), Immerwahrstrasse 2, 91058 Erlangen, Germany}

\author{Seyed Mehrdad Hosseini}
\affiliation{Heliatek GmbH, Dresden, Treidlerstraße 3, 01139 Dresden, Germany}

\author{Christian Uhrich}
\affiliation{Heliatek GmbH, Dresden, Treidlerstraße 3, 01139 Dresden, Germany}

\author{Christoph J. Brabec}
\affiliation{Department of Materials  Science and Engineering, Friedrich-Alexander-Universität Erlangen-Nürnberg, 91054 Erlangen, Germany}
\affiliation{Helmholtz Institute Erlangen-Nürnberg for Renewable Energy (HI ERN), Immerwahrstrasse 2, 91058 Erlangen, Germany}

\author{James R. Durrant}
\affiliation{Department of Chemistry and Centre for Processable Electronics"
MSRH, 82 Wood Lane, W12 0BZ}
\affiliation{SPECIFIC IKC, Department of Materials, University of Swansea, Bay Campus, Swansea, SA1 8EN, UK}

\author{Carsten Deibel}
\affiliation{Institut für Physik, Technische Universität Chemnitz, Straße der Nationen 62, D-09111 Chemnitz, Germany}

\author{Roderick C. I. MacKenzie}
\email[]{roderick.mackenzie@durham.ac.uk}

\affiliation{Department of Engineering, Durham University, Lower Mount Joy, South Road, Durham, DH1 3LE, UK}
\date{\today}
\begin{abstract}
Over the last two decades the organic solar cell community has synthesised tens of thousands of novel polymers and small molecules in the search for an optimum light harvesting material.  These materials were often crudely evaluated simply by measuring the current voltage curves in the light to obtain power conversion efficiencies (PCEs). Materials with low PCEs were quickly disregarded in the search for higher efficiencies. More complex measurements such as frequency/time domain characterisation that could explain why the material performed as it did were often not performed as they were too time consuming/complex. This limited feedback forced the field to advance using a more or less random walk of material development and has significantly slowed progress. Herein, we present a simple technique based on machine learning that can quickly and accurately extract recombination time constants and charge carrier mobilities as a function of light intensity simply from light/dark JV curves alone. This technique reduces the time to fully analyse a working cell from weeks to seconds and opens up the possibility of not only fully characterising new devices as they are fabricated, but also data mining historical data sets for promising materials the community has over looked.
\end{abstract}

\maketitle

\section{Introduction}
Over the last 22 years organic solar cell efficiencies have risen from 2.5\% in 2001 \cite{Saheen_2.5efficient_2001}
to over 19\% \cite{Liu2022PM6Y6record} today. Much of this increase in performance can be attributed to steady improvement in material systems\cite{li_polymer_2012,liu_recent_2022}.
The first reported cells relied on blends of MEH-PPV/P3HT and C60 fullerene derivatives \cite{Saheen_2.5efficient_2001, Dennler_basicstoapplicarion_2005}.
Later in the late 2000s low band gap polymers started to emerge with alternating copolymers of fluorene with Donor-Acceptor-Donor (D-A-D) segments such as PTPTB  with efficiencies around 10\% \cite{https://doi.org/10.1002/1616-3028(200108)11:4<255::AID-ADFM255>3.0.CO;2-I}. In the late 2010s the community moved away from fullerene based acceptors to small molecules, with this came efficiencies nearing 20\%\cite{Lui_replacefullerene_2013,zhang_material_2018, hou_organic_2018}.
Although efficiencies are slowly increasing at a rate of around 1\% a year it takes tremendous effort from thousands of researchers across the world to achieve this. Furthermore, quantities such as device life time and efficiency still need to be significantly optimised before commercialisation can be considered for polymer cells \cite{riede_commercialsuccess_2020, Duan_stability_2020}.
This points to another decade of slowly improving device performance that humanity can ill afford given the rapidly rising global temperatures \cite{intergovernmental_panel_on_climate_change_ipcc_climate_2023}.
Part of the reason for this slow progress in organic photo-voltaics (OPV) development is a lack of timely and detailed feedback to chemists from device engineers \cite{Yang_ternary_2013,Kranthiraja_experimentML_2021}.
Typically a new material will be synthesised and then used to fabricate a few test devices using a handful of solvents and a few annealing temperatures. Simple current-voltage (JV) curve sweeps will be performed to determine Power Conversion Efficiency (PCE), Fill Factor (FF), Open Circuit Voltage ($V_\mathrm{oc}$) and short-circuit current ($j_\mathrm{sc}$). These measurements will take only seconds and allow the scientist to see if the material has good photovoltaic properties. However, JV measurements will not give information as to why the device/material works well or poorly and do not give hints as to how material form/function should be improved.  To obtain this information one has to perform more time consuming measurements to extract key device parameters such as recombination rate, charge carrier mobility, and measures of disorder. Examples of techniques that can extract this information are, impedance spectroscopy (IS) \cite{macdonald_impedance_1992,Santiago_ISOPV_2011}, Impedance Modulated Photocurrent Spectroscopy (IMPS) \cite{DICARMINE20083744}, Impedance Modulated Photovoltage Spectroscopy (IMVS) \cite{Set_modelIMVS_2015,Byers_IMVSIMPSp3ht_2011}, Transient Photocurrent (TPC) \cite{Street_stateTPC_2011,Vollbrecht_chargecarrierdensity_2022}, Transient Photovoltage (TPV) \cite{Foertig_nongeminaterec_2012,Bisquert_fromfrequency_2021,Nakano_differentialcharge_2019} and charge extraction (CE) measurements \cite{Tress_chargeextraction_2013,Hanfland_meaningCELIV_2013}.

Although considerable efforts have gone into refining these methods they remain complex and require expertise and equipment that is often not found in the same lab as the people with knowledge in synthesis. Other approaches to get at fundamental device parameters such as fitting numerical models to experimental data can often take longer than the experiments themselves and also require expertise and models which are rarely found in the same place as where the material is fabricated \cite{Mackenzie2012extractingmicroscopic}.
Thus very often without detailed characterisation the scientist is left guessing as to why one molecule performs better than another or why devices fabricated under given conditions perform as they do. This makes it very difficult to determine the next steps in material/device optimisation.

Thus one can think of the development of OPV materials as a random walk, with chemists developing new materials and disregarding the majority of them as on first glance they do not perform.  Some more highly performing materials are occasionally investigated with more comprehensive methods (such as P3HT:PCBM in the past and more recently PM6:Y6).  This may well have led to promising materials being disregarded and skipped over as they did not perform well in the first batch or two of fabricated devices due to selecting the wrong solvents/annealing conditions or molecular weights. We are in effect searching for a needle in the hay stack but in the dark.

Although this problem is serious in the academic setting where a researcher may make a new material every few weeks, it is much worse in high through-put labs where new materials are generated daily. Candidate materials are often only tested against a few standard combinations of donor/acceptor molecules, solvents and annealed at a few temperatures before the materials are disregarded. Thus there exist a huge back catalogue of JV curves both in the literature and in the industry for material which were never fully analysed.

Our aim when writing this paper was to develop a method that can accurately extract charge carrier recombination time ($\tau$) and  mobility ($\mu$) as a function of light intensity using the most simple, quickest and easy to perform set of experiments possible.  We wanted a measurement technique that took seconds to apply, that anybody without expensive lasers/frequency domain equipment could use and enabled the feedback loop from device performance to material parameters to be efficiently closed for all in the community.   We focused on the recombination time constant and charge carrier mobility because they can be used to identify if recombination or transport is the key bottleneck in device performance, which can in turn give hints as to how to tune the molecular packing and/or morphology. Furthermore, when combined in the $\mu\cdot\tau$ product they give a standard benchmark for material performance \cite{Tumblestone_mutau_2012,Street_mutau_2012,Kirchhartz_mutau_2012}.

Herein, we demonstrate that both charge carrier recombination time  and  mobility can be extracted from JV curves alone using a combination of machine learning (ML) models trained on physically accurate device models.  We compare the values of recombination rate and charge carrier mobility extracted by our new method to values extracted by more traditional frequency domain/transient measurements from both spin coated and evaporated cells. Thus we develop a high throughput tool that has the potential to close the feedback loop and accelerate device development. 

\newpage
\section{Methods}
\subsection{Time domain measurements on evaporated devices}
\begin{figure}[H]
    \centering
    \includegraphics[width = 0.8\linewidth]{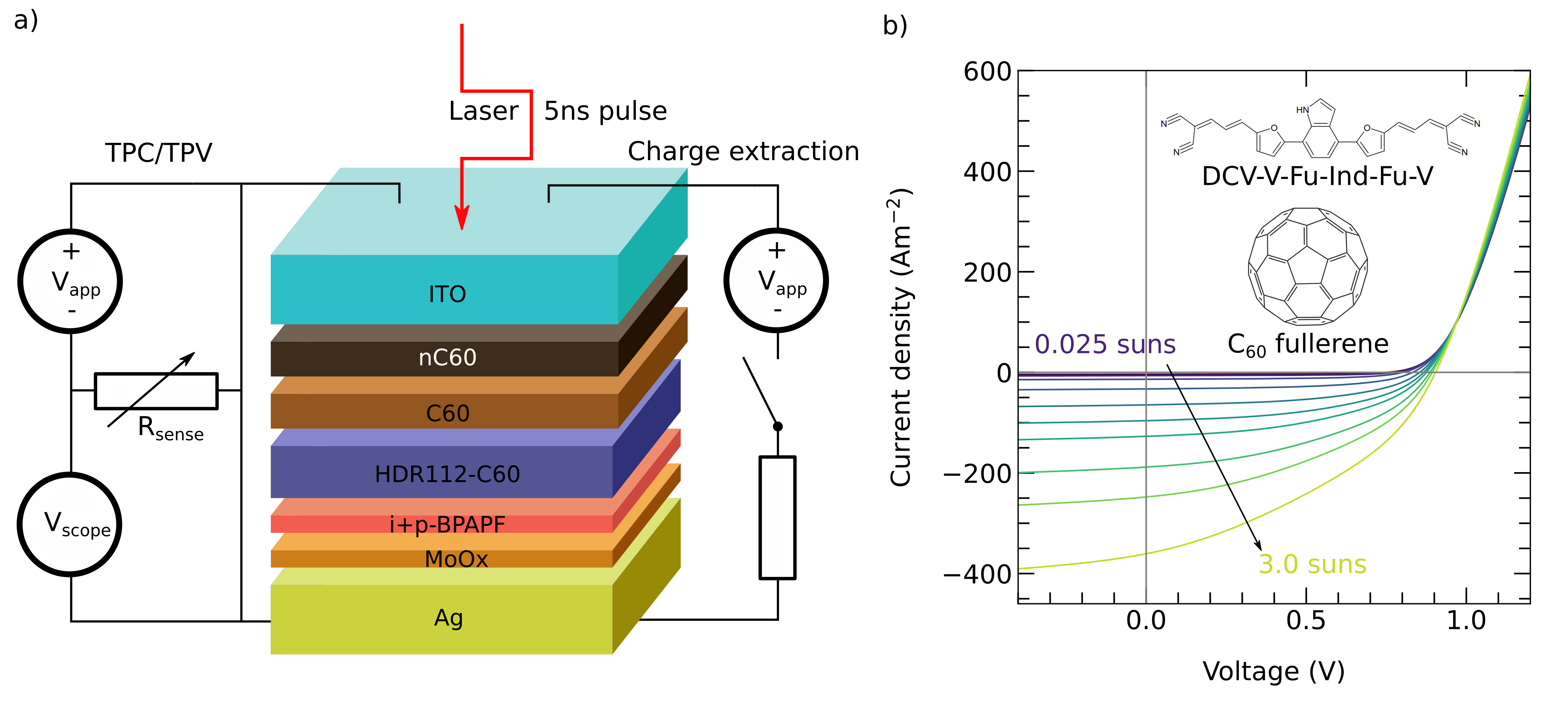}
    \caption{a) Device architecture and schematic depiction of transient techniques TPC/TPV and charge extraction ; b) Measured JV-curves from 0.025 Suns to 1 Suns for the device evaporated at room temperature. Inset: The molecular structures of DCV-V-Fu-Ind-Fu-V and C\textsubscript{60}.}
    \label{fig:sn21_overview}
\end{figure}

\begin{figure}[H]
    \centering
    \includegraphics[width = 0.35\linewidth]{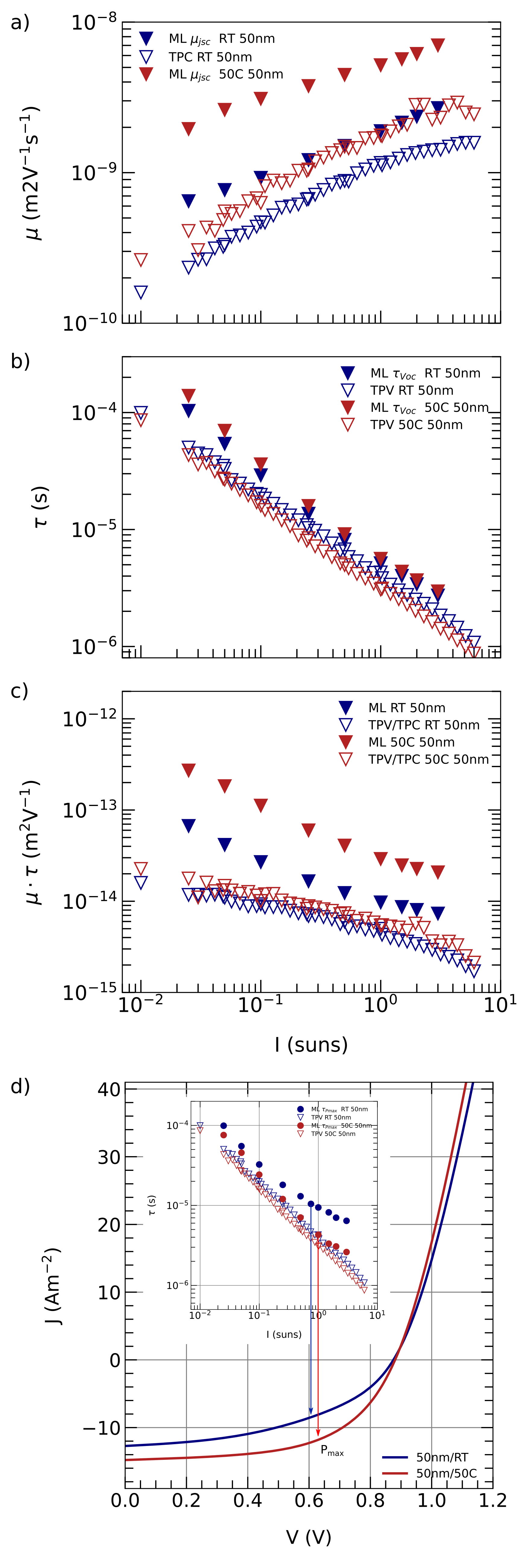}
    \caption{a) Light intensity dependent charge carrier mobility measured using charge carrier extraction for a device deposited at room temperature (blue)/$50~^{\circ}\mathrm{C}$ (red) ; b)  Light intensity dependent charge carrier lifetime measured using TPV for a device deposited at room temperature (blue)/$50~^{\circ}\mathrm{C}$ (red) ; c) The $\mu_\mathrm{jsc} \cdot \tau_\mathrm{Voc}$ product calculated from the above curves. In this figure the open triangles represent the experimental measurements and the solid triangles represent the results of the ML. d) JV-curves for devices deposited at room temperature (blue)/$50~^{\circ}\mathrm{C}$ (red); inset shows charge carrier lifetime, closed circles show the predicted lifetime at maximum power point P\textsubscript{max}. }
    \label{fig:sn21_predictions}
\end{figure}

Two devices of layer structure Glass/ITO/nC\textsubscript{60}/C\textsubscript{60}/DCV-V-Fu-Ind-Fu-V:C\textsubscript{60}/MoO\textsubscript{3}/Ag were deposited by evaporation, in one device the substrate temperature was held at 50$~^{\circ}\mathrm{C}$ during deposition of the active layer, while in the other device substrate temperature was allowed to float at room temperature \cite{Saladina_powerlaw_2023}. The device structure is depicted in Figure \ref{fig:sn21_overview}a while the molecular structures and example JV curves can be seen in Figure \ref{fig:sn21_overview}b. The active layer was 50 nm thick and made by co-evaporating the small molecule donor DCV-V-Fu-Ind-Fu-V with C\textsubscript{60}.  We performed TPV at open circuit and charge extraction at short circuit to measure recombination times and effective charge carrier mobility respectively. A summary of these measurements can be seen in Figure \ref{fig:sn21_predictions}.

Both JV curve and transient measurements were performed at light intensities ranging from 0.025 Suns to 3 Suns. It can be seen that the charge carrier mobility measured at $j_\mathrm{sc}$ is a factor two higher for the 50$~\mathrm{C}^{\circ}$ device than for the room temperature device. This is attributed to slightly better transport properties caused by favourable morphology. Lifetimes at $V_\mathrm{oc}$ are almost identical for both devices, indeed it can be seen from the JV-curves in Figure \ref{fig:sn21_predictions}d that $V_\mathrm{oc}$ is very close for both temperatures.

It is now our aim is to see if using the JV curves alone (see Figure \ref{fig:sn21_overview}b) coupled with machine learning we can predict all the data extracted using transient measurements presented in Figure \ref{fig:sn21_predictions}. JV curves are very quick and easy to measure. Thus if we were able to extract $\mu$ and $\tau$ from these curves alone months of measurement work could be saved. To do this we first set up the device structure in our drift-diffusion model OghmaNano \cite{Mackenzie2012extractingmicroscopic,oghma}.

The model solves Poisson's equation to take account of electrostatic effects within the device, electron/hole charge carrier continuity and drift-diffusion equations to describe carrier transport. Finally to describe carrier trapping and recombination, the LUMO and HOMO Urbach tails are each split up into 8 discrete trap levels and a Shockley-Read-Hall capture escape equation is solved for each energetic range. This approach allows carries to be described both in energy and position space within the device.  More detail about the model can be found elsewhere \cite{Mackenzie2011modelingp3htpcbm, Mackenzie2012extractingmicroscopic, Mackenzie2020ohmicspacecharge}.

\begin{figure}[H]
    \centering
    \includegraphics[width = 0.5\linewidth]{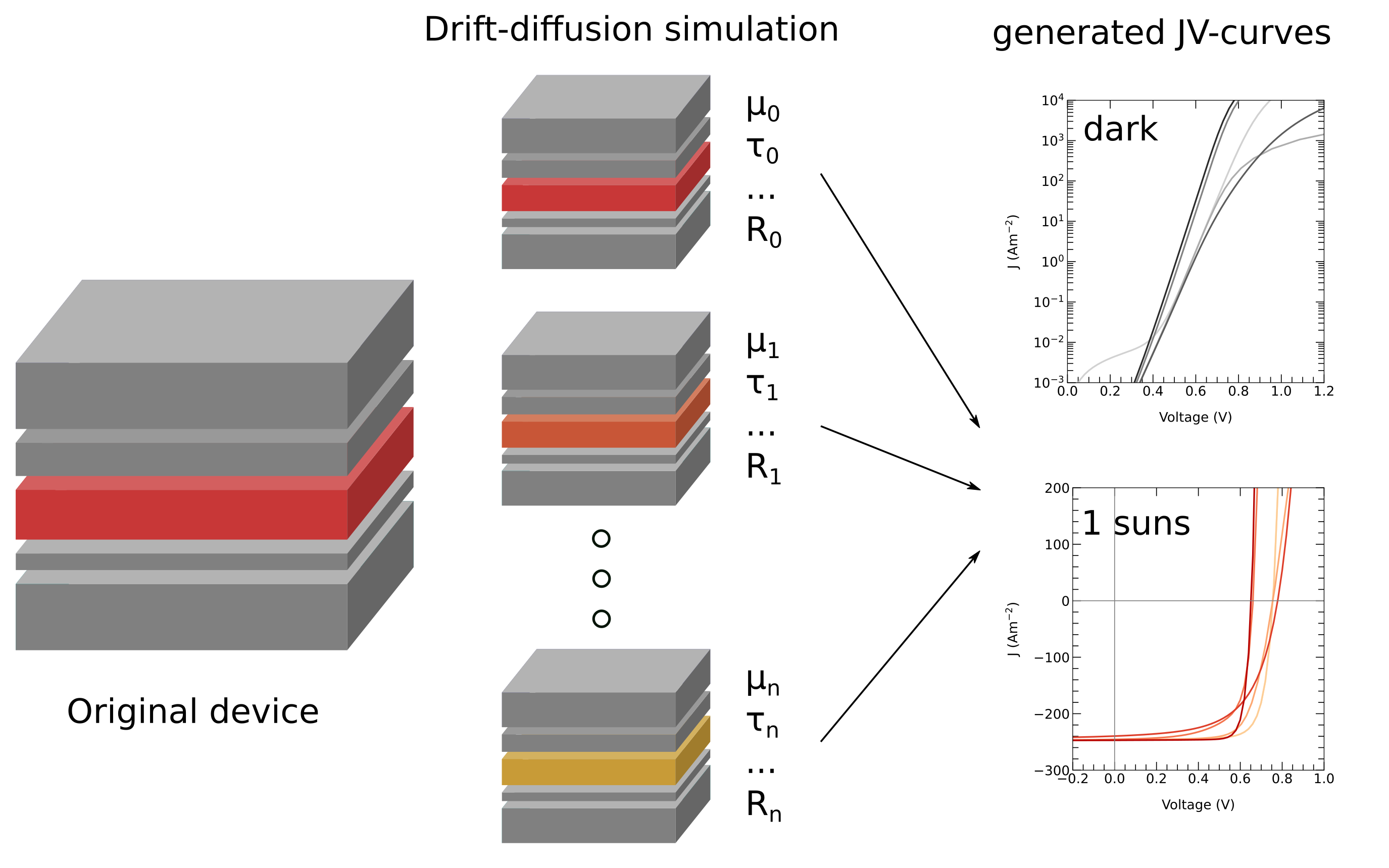}
    \caption{Creation of the training data set by artificially generating the device with randomly assigned parameters in a drift-diffusion simulation. The dark JV-curves and at $1~\mathrm{Suns}$ as well as recombination rate at $V_{oc}$ and mobility at $J_{sc}$ are simulated and stored.}
    \label{fig:ann_encoding}
\end{figure}

Using this base device structure, 20,000 copies of the simulation file were made to form a sample set of 20,000 virtual devices. Each virtual device had randomly assigned electron/hole mobilities, trap densities, Urbach tail slopes and carrier trapping/escape constants. From these devices 20,000 corresponding light and dark JV curves were generated. Furthermore, for each device the calculated recombination rate at $V_{oc}$ and charge carrier mobility at $J_{sc}$ were stored. This process is described in Figure \ref{fig:ann_encoding}.

Generating this data set takes around two hours and provides the basis for training the machine learning algorithm. The advantage of training the machine learning algorithm on virtual data is that most machine learning algorithms are very data hungry requiring thousands of examples to learn. Furthermore, it enables us to know exactly what the recombination rate is at $V_{oc}$ (mobility at $J_{sc}$) which would be hard to do experimentally.

\begin{figure}[H]
    \centering
    \includegraphics[width = 0.5\linewidth]{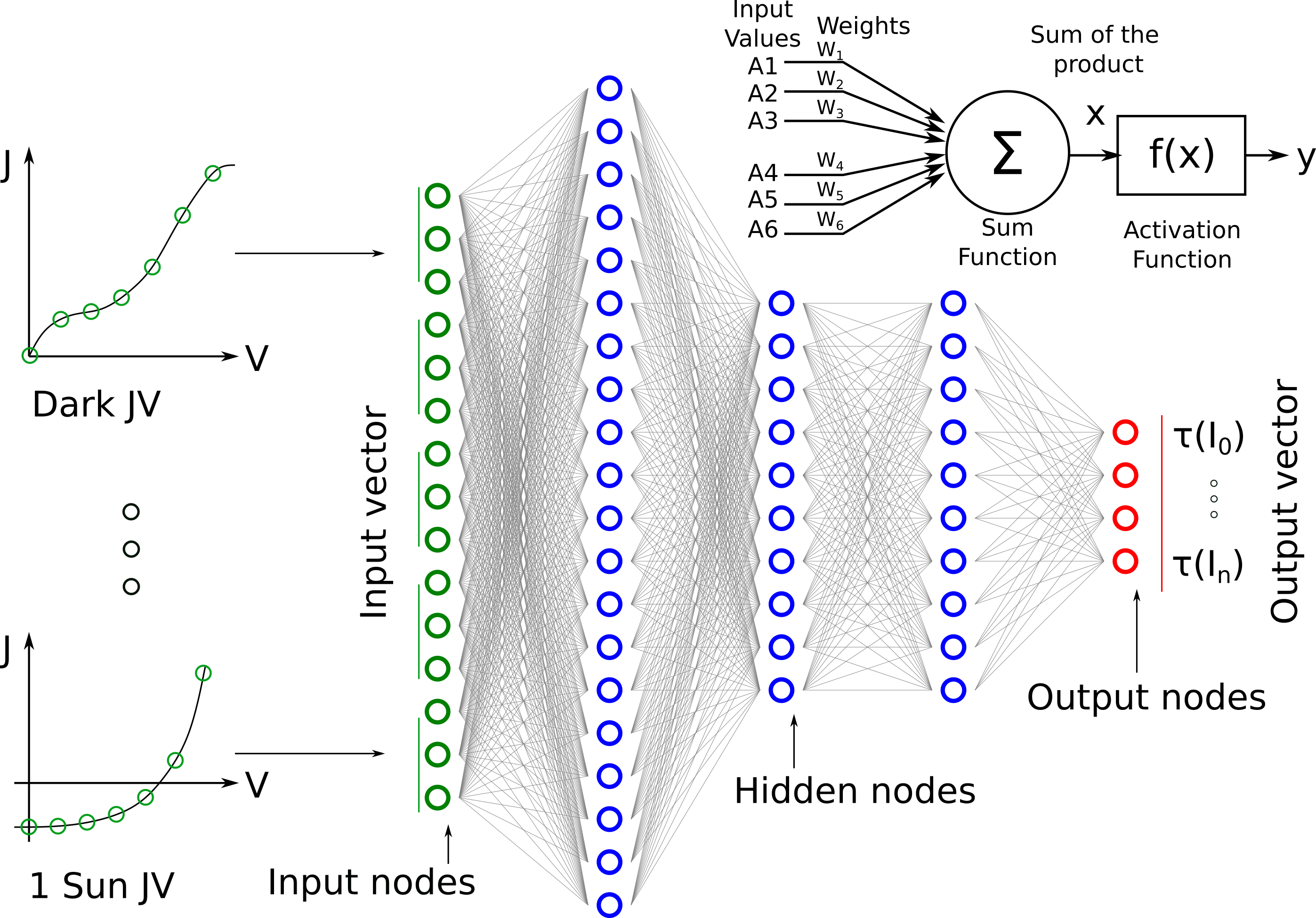}
    \caption{A diagram of the neural network used to extract material parameters from the data within this paper. Visible on the left hand side of the image is the experimental (or simulated) data, with the green dots on the curves representing the points at which the curves were sampled to form input vectors for the neural network. The JV-curves are being sampled at discrete voltages to provide data points to the neural networks input nodes. Any number or combination of experimental measurements can be placed on the input to the network, one simply has to extend the number of input neurons, and retrain the network. The neural network itself has green input nodes, blue hidden layers, and red output nodes. Each output node corresponds to a device/material parameter such as charge carrier mobility or recombination rate. Inset: A single neuron.}
    \label{fig:ann_idea}
\end{figure}

The next task is to train the machine learning algorithm with the data. This is depicted in Figure \ref{fig:ann_idea}.  For each device in turn the light and dark JV curves are presented to the inputs of the neural network. The network is then asked to predict the values of charge carrier mobility and recombination rate as a function of light intensity on the outputs. At the start of training the model predicts these values quite poorly, however as training progresses and the network sees more examples, the predicted values of $\mu$ and $\tau$ for each JV curve become closer to the correct values (more details on the training can be found in the SI). Once the network has been trained on all devices, the order of the devices are shuffled and training begins again, this process repeats until the network can correctly predict $\mu$ and $\tau$ for any given JV curve in the data set. Once the error is sufficiently small, the weights are fixed and the model is ready to predict on experimental data. To test the ability of the network to extract $\mu$ and $\tau$ from as of yet unseen data, 20\% of the 20,000 training set is kept out of the training process, and used at the end of the training process to assess the performance of the network. Once the model was trained on virtual data to our satisfaction, the experimental JV curves for each device in Figure \ref{fig:sn21_overview}b were fed into the neural network in an attempt to predict the values in Figure \ref{fig:sn21_predictions}.

The values of $\tau$ and $\mu$ predicted from the JV curves are shown in Figure \ref{fig:sn21_predictions} as solid triangles. It can be seen that the predicted values follow those of the directly measured values within one order of magnitude, accurately following the trend of the experimental data. This demonstrates that there is indeed enough information in the JV curves alone to determine $\tau$ and $\mu$. As $V_\mathrm{oc}$ is almost the same for both devices, the information gained with TPV is limited in our case. But the machine learning model enables to also predict lifetimes at the maximum power point $P_\mathrm{max}$. The inset in \ref{fig:sn21_predictions}b shows this prediction. As the maximum power point for the room temperature device is at a lower voltage, the charge carrier density may be lower than at the maximum power point for the $50~^{\circ}\mathrm{C}$ device and therefore result in longer carrier lifetime.

\subsection{Frequency domain measurements on spin coated devices}
\begin{figure}
    \centering
    \includegraphics[width = 0.8\linewidth]{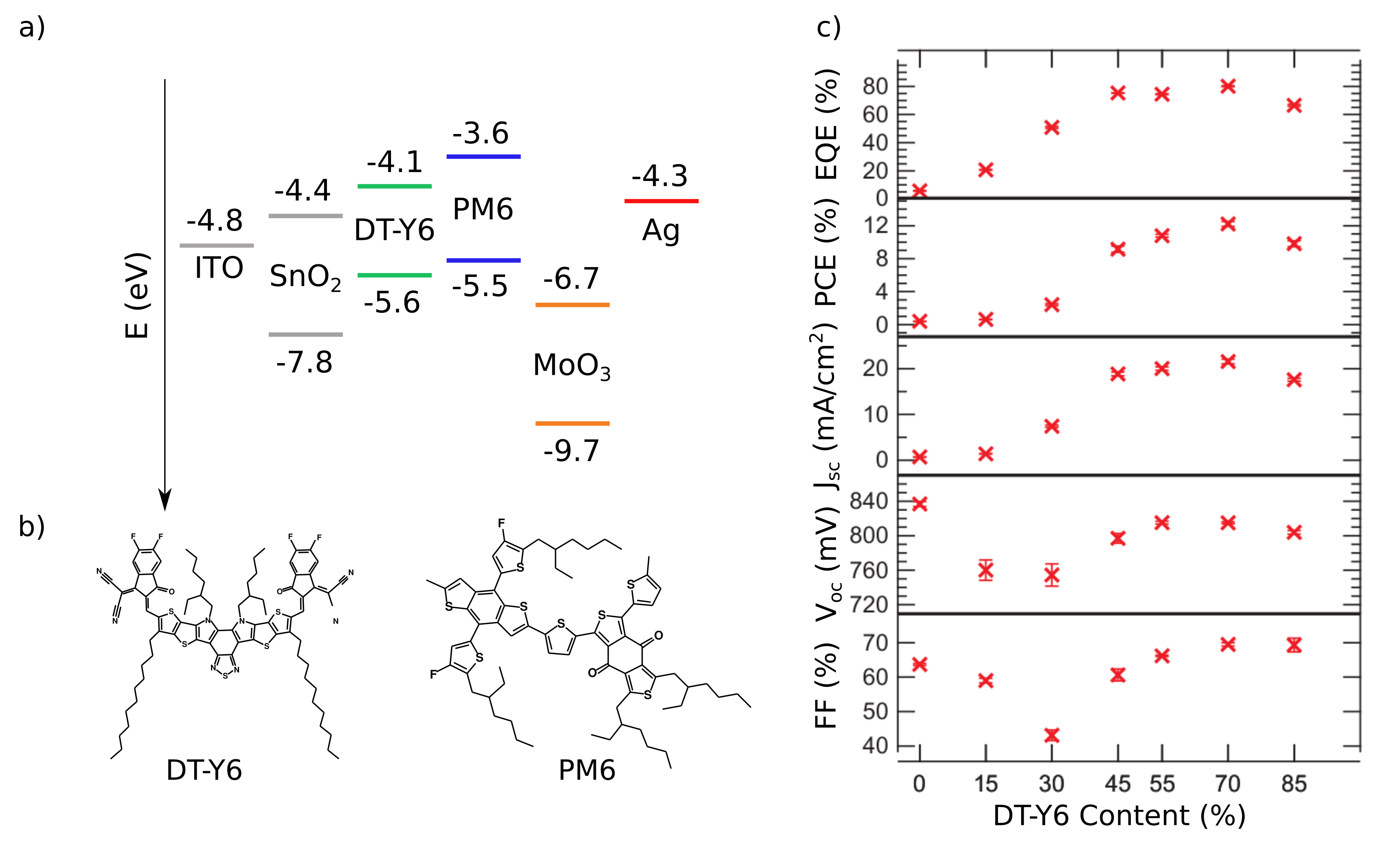}
    \caption{a) Device structure; b) Polymers of the active layer c) Device parameters depending on DT-Y6 content.}
    \label{fig:pm6dty6_overview}
\end{figure}

In the previous section we compared the ability of machine learning to extract $\tau$ or $\mu$ from JV curves to the values $\tau$ or $\mu$ extracted from transient measurements. In this section, we demonstrate the general ability of our ML-approach by turning our attention to state-of-the-art PM6:DT-Y6 spincoated devices measured using frequency domain techniques.
A series of glass/ITO/SnO\textsubscript{2}/PM6:DT-Y6/MoO\textsubscript{3}/Ag devices with varying DT-Y6 content were fabricated. The ratios chosen were 0:100, 15:85, 30:70, 45:55, 55:45, 70:30 and 85:15 of DT-Y6 to PM6 respectively. The molecular structure of these materials along with the device structure can be seen in Figure \ref{fig:pm6dty6_overview}a,b. Current voltage curves were measured under AM1.5G illumination to obtain $V_{oc}$, PCE, $J_{sc}$ and FF are plotted in Figure \ref{fig:pm6dty6_overview}c as a function of blend ratio (See SI for full curves). It can be seen that as the DT-Y6 ratio increases so does the PCE with a maximum PCE observed at around 70:30.

\begin{figure}
    \centering
    \includegraphics[width = 0.9\linewidth]{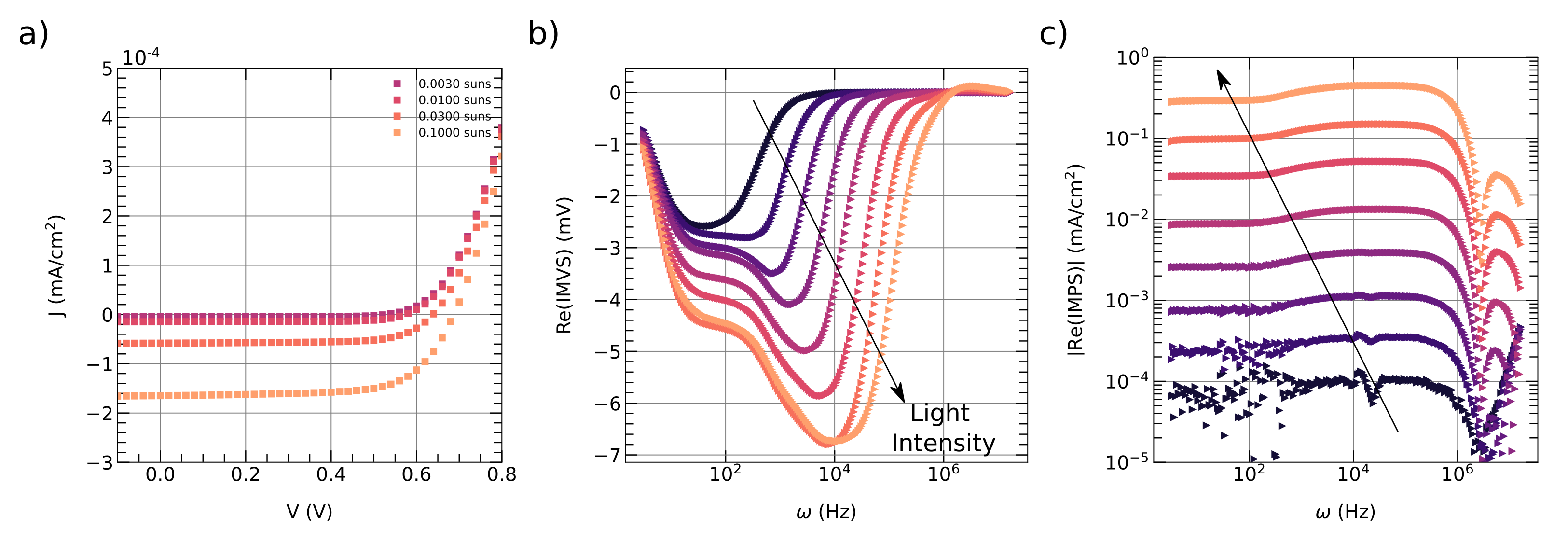}
    \caption{a) JV-curves for selected intensities; b) Intensity dependent experimental IMVS; and c) IMPS data for the  45\% DT-Y6 device.}
    \label{fig:pm6dty6_experiment}
\end{figure}

\begin{figure}
    \centering
    \includegraphics[width = \linewidth]{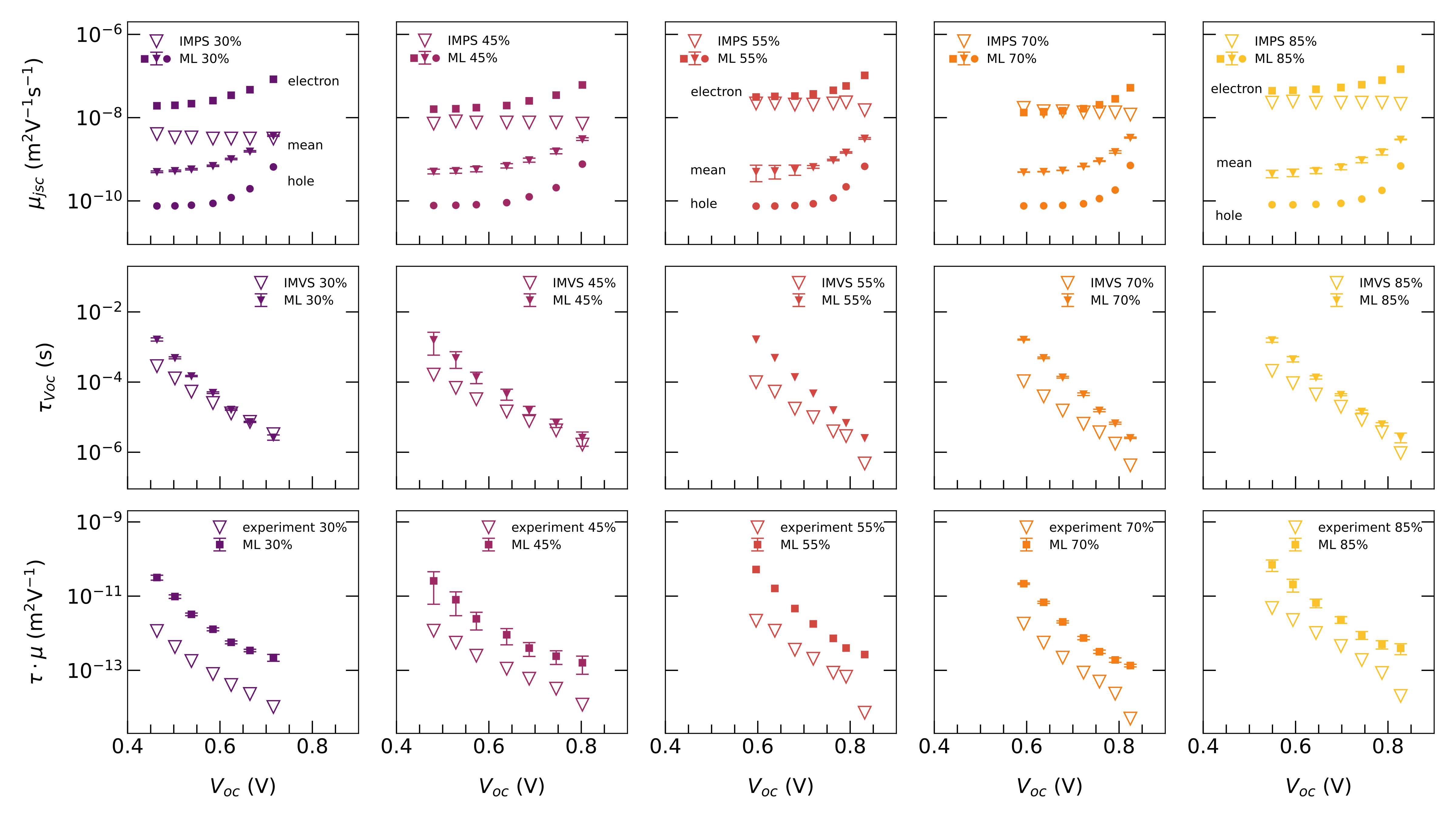}
    \caption{Top row) Open triangles represent charge carrier mobility as a function of light intensity measured at $J_{sc}$ using IMPS for varying DT-Y6 content. Middle row) Open triangles represent recombination time constants measured at $V_{oc}$ using IMVS plotted as a function of light intensity for varying DT-Y6 content. Bottom row) Open triangles represent calculated $\mu \cdot \tau$ products from the above two rows. Closed markers represent predicted values extracted from JV curves alone using machine learning. Closed squares relate to electrons, circles to holes and triangles to the (geometric) mean.}
    \label{fig:pm6dty6_predictions}

\end{figure}

To investigate the performance of these devices in terms of charge carrier transport and recombination rate we performed IMVS at open circuit and IMPS at short circuit to obtain charge carrier recombination rates and mobility as a function of light intensity. Example IMVS/IMPS curves are shown in Figure \ref{fig:pm6dty6_experiment}b,c from $300~\mu\mathrm{Suns}$ to $1~\mathrm{Suns}$. The experimental charge carrier lifetimes were calculated from the real part of the IMVS-signal and charge carrier mobilities were inferred from the real part of the IMPS-signal. A summary of these measurements can be seen in the top two rows of Figure \ref{fig:pm6dty6_predictions}. Note all data extracted from experiment is plotted as open triangles, the closed triangles are the results of the machine learning and will be discussed later. If the $\mu\cdot\tau$ product is examined (bottom line of the figure) it can be seen that $\mu\cdot\tau$ is higher for the high performing DT-Y6 ratios (55:45, 70:30 and 85:15) mainly due to a higher effective charge carrier mobility.

The above results represent a base line against which to compare the machine learning. Before we go further however, it is worth underlining some of the points made in the introduction about detailed characterisation being the bottleneck to device development by noting that the above measurements took around 6 months to measure and analyse by hand.

Again the experimental JV curves for each device in Figure \ref{fig:pm6dty6_experiment}a were fed into the neural network in an attempt to predict the values in Figure \ref{fig:pm6dty6_predictions}. The predicted values are shown as solid triangles for mean values (geometric mean in case of charge carrier mobility), solid squares for electrons and solid circles for holes.
Taking the top row of graphs first, it can be seen that the model predicts electron mobility to be orders of magnitude higher than hole mobility. This is in accordance with literature \cite{tokmoldin_extraordinarily_2020}. Further the predicted electron mobility is in good agreement with the experimental IMPS data. As the electrons are the faster charge carrier species they dominate the IMPS response. Due to their low charge carrier mobility holes will not be able to follow the high frequencies.
Examining the second line of graphs it can be seen that the Neural Network can predict the absolute value of the recombination time constant as a function of light intensity very well with the error being slightly higher for the lower light intensities. Still the error stays well below one order of magnitude. Furthermore the trend of the lifetime is also accurately reproduced. 
The bottom row of graphs compare the predicted $\mu_\mathrm{jsc,e}\cdot\tau_\mathrm{Voc}$ product to the measured values with these trends also agreeing well.

Finally, it should be noted that the error bars in Figure \ref{fig:pm6dty6_predictions} on the ML results were generated using a second Neural Network acting as an error estimation/confidence network. 
We used the 20\% fraction of the training set that the $\mu$ or $\tau$ predicting network had not been exposed to, to train the error estimation network. The learning procedure was to ask our $\mu$ or $\tau$ neural network to guess $\tau$ and $\mu$ for a JV curve it had not yet seen. We would then ask our error estimation Network to predict the expected error in the guess of $\tau$ and $\mu$. The error estimation network was then iteratively trained to try to improve its understanding of how good the values of $\mu/\tau$ would be for a given JV curve. As is visible in Figure \ref{fig:pm6dty6_predictions} the error prediction network is fairly confident about the ability of the $\mu/\tau$ to be predictive. This error however should not be treated as an absolute measure of accuracy but treated as a flag to determine if the experimental JV curve is far from something the $\mu/\tau$ has had experience with.

\section{Discussion}
Above we have demonstrated that using a combination of ML algorithms trained on simulated JV curves alone, one can build a tool to extract charge carrier mobility and recombination rate as a function of light intensity, thus removing the need for time consuming and costly characterisation.  We anticipate this tool being used by the community to quickly screen new devices and materials and also as a tool to screen the vast historical data sets available in the literature and in industry. The method can also be thought of as a tool to democratise the characterisation of OPV devices. Currently only well funded labs can perform mobility and life time measurements as they require relatively expensive lasers. This tool will allow more people to start extracting this data.

In some ways it is remarkable that using a simple drift diffusion model and a machine learning algorithm we are able to extract carrier recombination time and charge carrier mobility as a function of light intensity. One would have though that some type of transient measurement was needed to extract this information. However, this preconception comes from a human centric view of solar cell measurements, in that one thinks measurements such as TPC and SCLC are needed to measure charge carrier mobility because that is what has been done in the past.  However, we should approach the problem from the perspective of Shannon entropy. Entropy in information theory
\cite{shannon_entropie_1948}
is a measure of how much information is in a signal. For example a photograph of a perfectly clear blue sky contains low entropy (embodied information) as it simply tells you it is a sunny day. However a picture of a clouded sky has higher entropy (embodied information), as it can tell you how high the clouds are, what type of clouds there are, likelihood of rain and likelihood of thunder.  We should therefore think of electrical/optical measurements in the same context and ask how much embodied information does the measurement signal contain? In this case it is clear JV curves do encode information about $\tau$ and $\mu$ that the Neural Network can find and decode.

Continuing this line of reasoning, there is no reason why we should focus our efforts on decoding JV curves or other standard measurements such as TPC alone.  There may be another, as of yet unknown, measurement that may be as easy to obtain as a JV curve but contain more information that a machine learning algorithm can extract.  In other words, an experiment designed for machine learning extraction rather than for human extraction. Indeed, it may be that the machine has to design it's own perfect experiment to extract maximum possible information from a solar cell.

Now we comment on accuracy, although we demonstrated above that our method is accurate for the devices we chose. It should also be noted that it does not need to be completely accurate for all unusual classes of devices to be successful. Our method just needs to be good enough to show trends between devices and also flag up promising materials which are unusual. This first sift can then be used to flag devices to be investigated with more traditional experimental methods. 

A general comment should be made about the measurement of $\tau$ and $\mu$. It should be noted that the fundamentally difficult thing about measuring $\tau$ and $\mu$ in organic devices is that they are both a very strong function of carrier density due to the large number of trap states in the materials. Thus if applied voltage, photon flux, or contact materials are changed  $\tau$ and $\mu$ will change.  Therefore it is well known that different experiments that subject a device to different experimental conditions will produce different values of mobility/lifetime. For example both Charge Extraction by Lineally Increasing Voltage (CELIV) and TPC are commonly used to measure charge carrier mobility. In CELIV the device is held at $V_{bi}$ under constant illumination and a negative voltage ramp is applied to study charge carrier mobility while in TPC the device is usually held at $J_{sc}$ and the response of the device to a laser pulse is used to calculate mobility. Generally such measurements will produce values of mobility within an order of magnitude to each other with trends that agree but will not be identical. Thus it should be noted that when we compare our simulated values to the experimental values we are not comparing identical quantities (as it always is the case in organics). Our simulated values of $\tau$ and $\mu$ are defined as:

\begin{equation}\label{eq:paper_mueff}
    \mux{eff} = \frac{1}{d} \int\limits_{0}^{d}  \frac{\mux{free} n_\mathrm{free}(x)}{n_\mathrm{free}(x) + n_\mathrm{trap}(x)} \mathrm{d}x
\end{equation}
where $\mux{free}$ is the charge carrier mobility of completely free carriers, $n_\mathrm{free}$ is the density of completely free carriers and $n_\mathrm{trap}$ is the density of trapped carriers. The effective mobility is calculated for each charge carrier specimen separately and an average mobility is calculated by taking the geometric mean:

\begin{equation}
    \mu = \sqrt{\mux{e}\cdot\mux{h}}~.
\end{equation}

The lifetime $\tau$ is calculated by:

\begin{equation}
    \tau = \frac{\left(n_\mathrm{total} - n_\mathrm{0}\right) \cdot \left(p_\mathrm{total} - p_\mathrm{0}\right)}{R}
\end{equation}
with $n,p_\mathrm{total}$ being the total charge carrier density in the device, $n_\mathrm{0}$ the equilibrium free charge carrier density and $R$ the total recombination rate.

Thus some of the error in the graphs may be down to slightly different definitions of mobility and time constant.
Further it has been shown that charge carrier mobility results for the same device vary up to one order of magnitude when using different measurement techniques and up to a factor of three when different scientist analyse an identical dataset \cite{blakesley_towards_2014}. Difference between the ML predictions and experimental measurements are within the expected experimental error.

\begin{figure}[H]
    \centering
    \includegraphics[width = 0.5\linewidth]{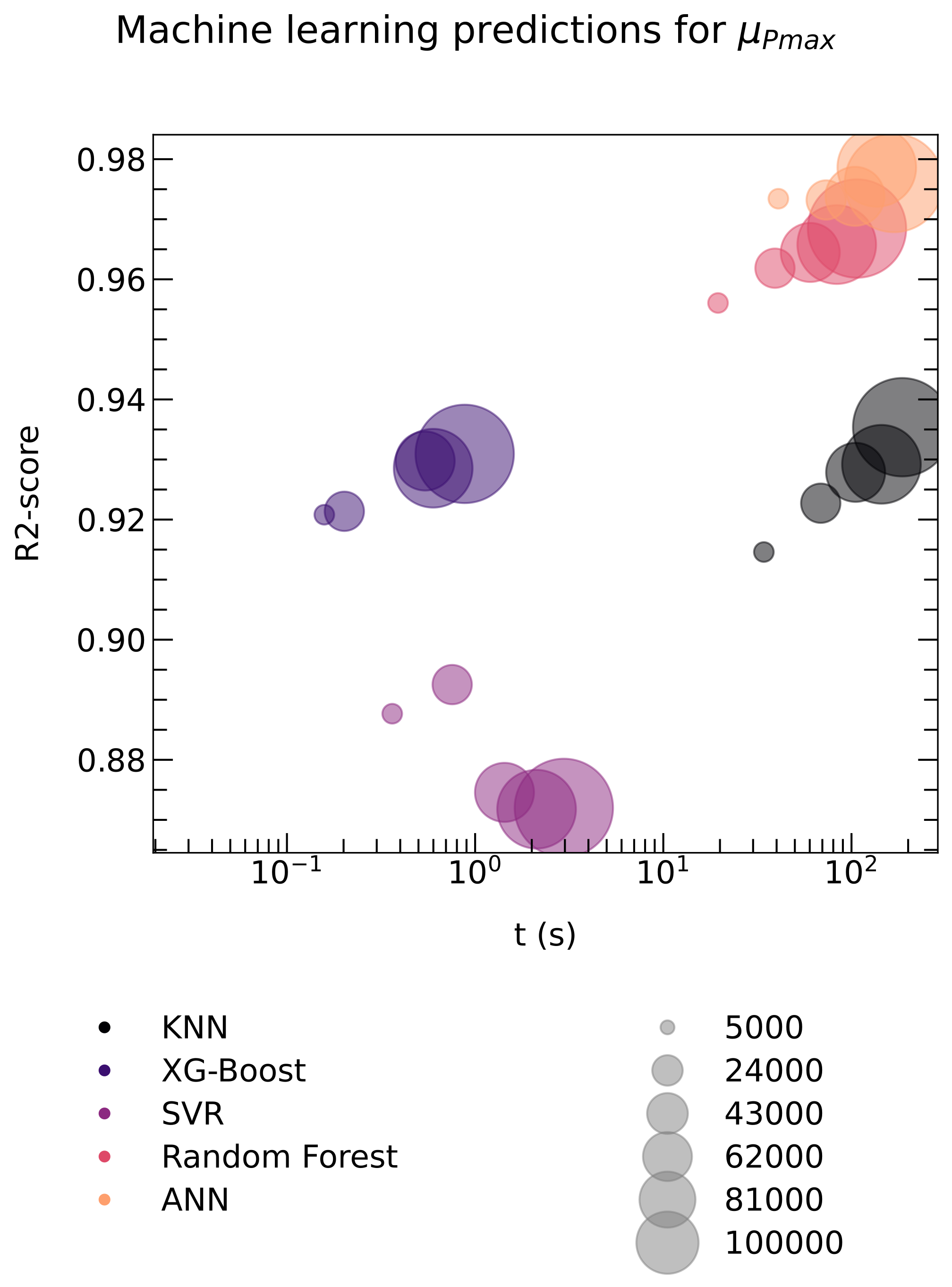}
    \caption{Comparison of accuracy and time taken to train Neural Networks,  k-nearest neighbour regression (KNN), random-forest regression, extreme-boosted-gradient-descent regression (XG-Boost) and support-vector regression (SVR) on the SN21 data set. It can be seen the Neural Network performs best but is slowest to train.}
    \label{fig:ml_methods}
\end{figure}

Finally, in the above examples we used Neural Networks for the machine learning, this is because we found their performance to be more accurate than other more traditional methods. Neural Networks do however require a lot of data and are also relatively slow to train. For comparison figure \ref{fig:ml_methods} plots the machine learning results from four other methods these include, k-nearest neighbour regression (KNN),
\cite{Cover_Knn_1967}
random-forest regression,
\cite{breiman_random_2001}
extreme-boosted-gradient-descent regression (XG-Boost)
\cite{Chen_XGboost_2016}
and support-vector regression (SVR) \cite{drucker_support_1996}.
The figure plots $R2$ score (accuracy) v.s. time taken to train for the data set generated for the PM6:DT-Y6 device. The size of the bubble represent the size of the training data set. Data sets of between 5000 and 100,000 devices were used. It can be seen that the XG-Boost algorithm is the fastest but also the worst, SVRs and KNNs have the same level of performance while  KNN is slower. The best performing method is the Neural Network, closely followed by the random forest. Each of these algorithms can be optimised, for example the number and size of layers in the Neural Network can be tuned to obtain best performance. However, these results represent our best efforts.

\section{Predicting on databases}

\begin{figure}[H] 
    \centering
    \includegraphics[width = \linewidth]{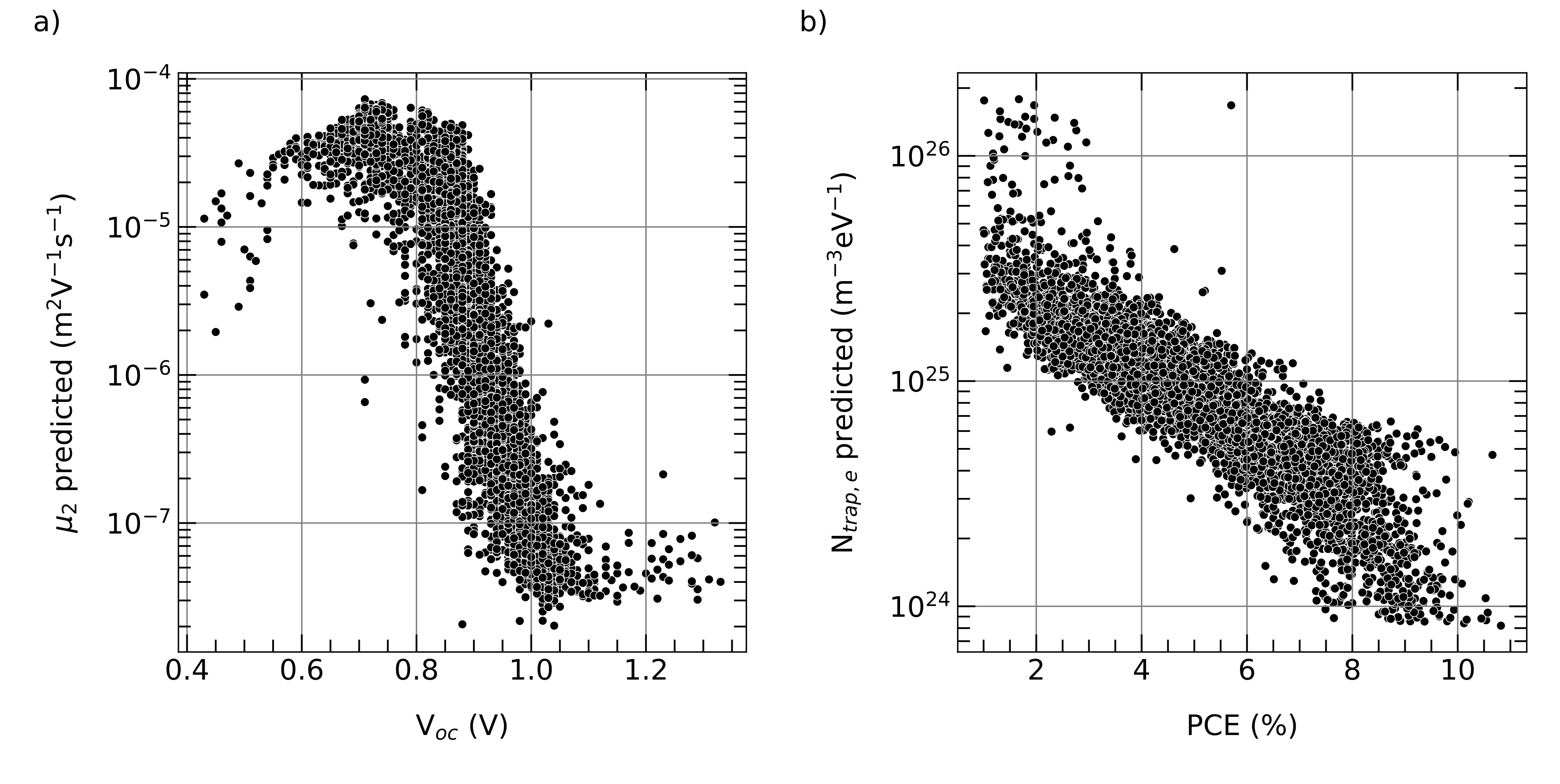}
    \caption{Predicted device parameters of a database containing around 10000 devices. The predictions are plotted over the experimentally determined $V_\mathrm{oc}$ or PCE. The colour code distinguishes planar- and bulk-hetero-junction devices. a) Mobility at $V_\mathrm{oc}$ b) trap state density for electrons.}
    \label{fig:db_predictions}
\end{figure}

The real strength of the machine learning approach is revealed when large sets of data have to be analysed, as it enables material parameters to be extracted that have not directly been measured. Indeed, the devices may have been made and discarded years ago.  As a demonstration of our method the ML algorithm was used to predict mobility and trap state density from a set of over 10000 historical JV curves held by Heliatek GmbH, the results can be seen in Figure \ref{fig:db_predictions}. The original database only contained JV-curves at dark conditions and at $1~\mathrm{Suns}$ light intensity. It can be seen that the model identifies a clear correlation between $V_{oc}$ and charge carrier mobility, as well as a clear correlation between PCE and trap density.

This technique would allow one to data mine these historical data sets and identify devices with optimal charge carrier transport properties that were potentially overlooked in the past.

\section{Conclusion}
Above we demonstrated that one does not need complex time domain/frequency domain measurement techniques to access charge carrier mobilities and recombination time constants.  This information is encoded within the far more simple to obtain current voltage curves. One simply needs a relatively low cost computer to extract this information. Furthermore, once trained the machine learning models take a fraction of a second to apply which means devices can be analysed as they are produced. This is important in the academic setting but more important in an industrial setting where tens of devices are produced per day. Furthermore, this approach will allow researchers to scour historical materials for promising candidates that we have skipped over as a community. Finally we emphasise that experimental data should be seen from an information theory point of view. Maximising entropy by conducting the right combination of experiments will be key to optimise the use of machine learning.

\begin{acknowledgments}
We thank Heliatek GmbH for funding MH's PhD through the EPSRC Centre for Doctoral Training in Renewable Energy Northeast Universities (ReNU). We thank the Deutsche Forschungsgemeinschaft (DFG Research Unit FOR 5387 POPULAR, Project No. 461909888) for their support.  RAP would further like to acknowledge the support through the Centre for Processable Electronics CDT program as well as the ATIP project (grant number EP/L016702/1 and EP/T028513/1). 
This work was supported by the Engineering and Physical Sciences Research Council (grant number EP/S023836/1).
\end{acknowledgments}

\bibliography{main}

\providecommand{\noopsort}[1]{}\providecommand{\singleletter}[1]{#1}%
\begin{thebibliography}{10}

\bibitem{Saheen_2.5efficient_2001}
S.~E. Shaheen, C.~J. Brabec, N.~S. Sariciftci, F.~Padinger, T.~Fromherz, and
  J.~C. Hummelen, ``{2.5\\
  Applied Physics Letters}, vol.~78, p.~841, 02 2001.

\bibitem{Liu2022PM6Y6record}
L.~Zhu, M.~Zhang, J.~Xu, and et~al., ``Single-junction organic solar cells with
  over 19\% efficiency enabled by a refined double-fibril network morphology,''
  {\em Nat. Mater.}, vol.~21, p.~656, 2022.

\bibitem{li_polymer_2012}
G.~Li, R.~Zhu, and Y.~Yang, ``Polymer solar cells,'' {\em Nature Photonics},
  vol.~6, p.~153, Mar. 2012.
\newblock Number: 3 Publisher: Nature Publishing Group.

\bibitem{liu_recent_2022}
Y.~Liu, B.~Liu, C.-Q. Ma, F.~Huang, G.~Feng, H.~Chen, J.~Hou, L.~Yan, Q.~Wei,
  Q.~Luo, Q.~Bao, W.~Ma, W.~Liu, W.~Li, X.~Wan, X.~Hu, Y.~Han, Y.~Li, Y.~Zhou,
  Y.~Zou, Y.~Chen, Y.~Li, Y.~Chen, Z.~Tang, Z.~Hu, Z.-G. Zhang, and Z.~Bo,
  ``Recent progress in organic solar cells ({Part} {I} material science),''
  {\em Science China Chemistry}, vol.~65, p.~224, Feb. 2022.

\bibitem{Dennler_basicstoapplicarion_2005}
G.~Dennler and N.~Sariciftci, ``Flexible conjugated polymer-based plastic solar
  cells: From basics to applications,'' {\em Proceedings of the IEEE}, vol.~93,
  no.~8, p.~1429, 2005.

\bibitem{https://doi.org/10.1002/1616-3028(200108)11:4<255::AID-ADFM255>3.0.CO;2-I}
A.~Dhanabalan, J.~K.~J. van Duren, P.~A. van Hal, J.~L.~J. van Dongen, and
  R.~A.~J. Janssen, ``Synthesis and characterization of a low bandgap
  conjugated polymer for bulk heterojunction photovoltaic cells,'' {\em
  Advanced Functional Materials}, vol.~11, no.~4, p.~255, 2001.

\bibitem{Lui_replacefullerene_2013}
T.~Liu and A.~Troisi, ``What makes fullerene acceptors special as electron
  acceptors in organic solar cells and how to replace them,'' {\em Advanced
  Materials}, vol.~25, no.~7, p.~1038, 2013.

\bibitem{zhang_material_2018}
J.~Zhang, H.~S. Tan, X.~Guo, A.~Facchetti, and H.~Yan, ``Material insights and
  challenges for non-fullerene organic solar cells based on small molecular
  acceptors,'' {\em Nature Energy}, vol.~3, p.~720, Sept. 2018.
\newblock Number: 9 Publisher: Nature Publishing Group.

\bibitem{hou_organic_2018}
J.~Hou, O.~Inganäs, R.~H. Friend, and F.~Gao, ``Organic solar cells based on
  non-fullerene acceptors,'' {\em Nature Materials}, vol.~17, p.~119, Feb.
  2018.
\newblock Number: 2 Publisher: Nature Publishing Group.

\bibitem{riede_commercialsuccess_2020}
M.~Riede, D.~Spoltore, and K.~Leo, ``Organic solar cells—the path to
  commercial success,'' {\em Advanced Energy Materials}, vol.~11, no.~1,
  p.~2002653, 2021.

\bibitem{Duan_stability_2020}
L.~Duan and A.~Uddin, ``Progress in stability of organic solar cells,'' {\em
  Advanced Science}, vol.~7, no.~11, p.~1903259, 2020.

\bibitem{intergovernmental_panel_on_climate_change_ipcc_climate_2023}
{Intergovernmental Panel On Climate Change (Ipcc)}, {\em Climate {Change} 2022
  – {Impacts}, {Adaptation} and {Vulnerability}: {Working} {Group} {II}
  {Contribution} to the {Sixth} {Assessment} {Report} of the
  {Intergovernmental} {Panel} on {Climate} {Change}}.
\newblock Cambridge University Press, 1~ed., 06 2023.

\bibitem{Yang_ternary_2013}
L.~Yang, L.~Yan, and W.~You, ``Organic solar cells beyond one pair of
  donor–acceptor: Ternary blends and more,'' {\em The Journal of Physical
  Chemistry Letters}, vol.~4, no.~11, p.~1802, 2013.
\newblock PMID: 26283112.

\bibitem{Kranthiraja_experimentML_2021}
K.~Kranthiraja and A.~Saeki, ``Experiment-oriented machine learning of
  polymer:non-fullerene organic solar cells,'' {\em Advanced Functional
  Materials}, vol.~31, no.~23, p.~2011168, 2021.

\bibitem{macdonald_impedance_1992}
J.~R. Macdonald, ``Impedance spectroscopy,'' {\em Annals of Biomedical
  Engineering}, vol.~20, p.~289, May 1992.

\bibitem{Santiago_ISOPV_2011}
F.~Fabregat-Santiago, G.~Garcia-Belmonte, I.~Mora-Seró, and J.~Bisquert,
  ``Characterization of nanostructured hybrid and organic solar cells by
  impedance spectroscopy,'' {\em Phys. Chem. Chem. Phys.}, vol.~13,
  pp.~9083--9118, 2011.

\bibitem{DICARMINE20083744}
P.~DiCarmine and O.~Semenikhin, ``Intensity modulated photocurrent spectroscopy
  (imps) of solid-state polybithiophene-based solar cells,'' {\em
  Electrochimica Acta}, vol.~53, no.~11, p.~3744, 2008.
\newblock Electrochemistry of Electroactive Materials Nanostructured
  Electroactive Materials Biological and Molecular Applications.

\bibitem{Set_modelIMVS_2015}
Y.~T. Set, B.~Li, F.~J. Lim, E.~Birgersson, and J.~Luther, ``{Analytical
  modeling of intensity-modulated photovoltage spectroscopic responses of
  organic bulk-heterojunction solar cells},'' {\em Applied Physics Letters},
  vol.~107, p.~173301, 10 2015.

\bibitem{Byers_IMVSIMPSp3ht_2011}
J.~C. Byers, S.~Ballantyne, K.~Rodionov, A.~Mann, and O.~A. Semenikhin,
  ``Mechanism of recombination losses in bulk heterojunction p3ht:pcbm solar
  cells studied using intensity modulated photocurrent spectroscopy,'' {\em ACS
  Applied Materials \& Interfaces}, vol.~3, no.~2, p.~392, 2011.
\newblock PMID: 21299191.

\bibitem{Street_stateTPC_2011}
R.~A. Street, ``Localized state distribution and its effect on recombination in
  organic solar cells,'' {\em Phys. Rev. B}, vol.~84, p.~075208, Aug 2011.

\bibitem{Vollbrecht_chargecarrierdensity_2022}
J.~Vollbrecht, N.~Tokmoldin, B.~Sun, V.~V. Brus, S.~Shoaee, and D.~Neher,
  ``{Determination of the charge carrier density in organic solar cells: A
  tutorial},'' {\em Journal of Applied Physics}, vol.~131, p.~221101, 06 2022.

\bibitem{Foertig_nongeminaterec_2012}
A.~Foertig, A.~Wagenpfahl, T.~Gerbich, D.~Cheyns, V.~Dyakonov, and C.~Deibel,
  ``Nongeminate recombination in planar and bulk heterojunction organic solar
  cells,'' {\em Advanced Energy Materials}, vol.~2, no.~12, p.~1483, 2012.

\bibitem{Bisquert_fromfrequency_2021}
J.~Bisquert and M.~Janssen, ``From frequency domain to time transient methods
  for halide perovskite solar cells: The connections of imps, imvs, tpc, and
  tpv,'' {\em The Journal of Physical Chemistry Letters}, vol.~12, no.~33,
  pp.~7964--7971, 2021.
\newblock PMID: 34388001.

\bibitem{Nakano_differentialcharge_2019}
K.~Nakano, Y.~Chen, and K.~Tajima, ``{Quantifying charge carrier density in
  organic solar cells by differential charging techniques},'' {\em AIP
  Advances}, vol.~9, p.~125205, 12 2019.

\bibitem{Tress_chargeextraction_2013}
W.~Tress, S.~Corvers, K.~Leo, and M.~Riede, ``Investigation of driving forces
  for charge extraction in organic solar cells: Transient photocurrent
  measurements on solar cells showing s-shaped current–voltage
  characteristics,'' {\em Advanced Energy Materials}, vol.~3, no.~7, p.~873,
  2013.

\bibitem{Hanfland_meaningCELIV_2013}
R.~Hanfland, M.~A. Fischer, W.~Brütting, U.~Würfel, and R.~C.~I. MacKenzie,
  ``{The physical meaning of charge extraction by linearly increasing voltage
  transients from organic solar cells},'' {\em Applied Physics Letters},
  vol.~103, p.~063904, 08 2013.

\bibitem{Mackenzie2012extractingmicroscopic}
R.~C.~I. MacKenzie, C.~G. Shuttle, M.~L. Chabinyc, and J.~Nelson, ``Extracting
  microscopic device parameters from transient photocurrent measurements of
  p3ht:pcbm solar cells,'' {\em Advanced Energy Materials}, vol.~2, no.~6,
  p.~662, 2012.

\bibitem{Tumblestone_mutau_2012}
J.~R. Tumbleston, Y.~Liu, E.~T. Samulski, and R.~Lopez, ``Interplay between
  bimolecular recombination and carrier transport distances in bulk
  heterojunction organic solar cells,'' {\em Advanced Energy Materials},
  vol.~2, no.~4, p.~477, 2012.

\bibitem{Street_mutau_2012}
R.~A. Street, A.~Krakaris, and S.~R. Cowan, ``Recombination through different
  types of localized states in organic solar cells,'' {\em Advanced Functional
  Materials}, vol.~22, no.~21, pp.~4608--4619, 2012.

\bibitem{Kirchhartz_mutau_2012}
T.~Kirchartz, T.~Agostinelli, M.~Campoy-Quiles, W.~Gong, and J.~Nelson,
  ``Understanding the thickness-dependent performance of organic bulk
  heterojunction solar cells: The influence of mobility, lifetime, and space
  charge,'' {\em The Journal of Physical Chemistry Letters}, vol.~3, no.~23,
  p.~3470, 2012.
\newblock PMID: 26290974.

\bibitem{Saladina_powerlaw_2023}
M.~Saladina, C.~W\"opke, C.~G\"ohler, I.~Ramirez, O.~Gerdes, C.~Liu, N.~Li,
  T.~Heum\"uller, C.~J. Brabec, K.~Walzer, M.~Pfeiffer, and C.~Deibel,
  ``Power-law density of states in organic solar cells revealed by the
  open-circuit voltage dependence of the ideality factor,'' {\em Phys. Rev.
  Lett.}, vol.~130, p.~236403, Jun 2023.

\bibitem{oghma}
R.~C.~I. MacKenzie, ``{OghmaNano}.'' \url{https://www.oghma-nano.com/}.
\newblock Accessed: 2010-09-30.

\bibitem{Mackenzie2011modelingp3htpcbm}
R.~C.~I. MacKenzie, T.~Kirchartz, G.~F.~A. Dibb, and J.~Nelson, ``Modeling
  nongeminate recombination in p3ht:pcbm solar cells,'' {\em The Journal of
  Physical Chemistry C}, vol.~115, no.~19, pp.~9806--9813, 2011.

\bibitem{Mackenzie2020ohmicspacecharge}
J.~A. Röhr and R.~C.~I. MacKenzie, ``Analytical description of mixed ohmic and
  space-charge-limited conduction in single-carrier devices,'' {\em Journal of
  Applied Physics}, vol.~128, no.~16, p.~165701, 2020.

\bibitem{tokmoldin_extraordinarily_2020}
N.~Tokmoldin, S.~Mehrdad~Hosseini, M.~Raoufi, L.~Quang~Phuong, O.~J.~Sandberg,
  H.~Guan, Y.~Zou, D.~Neher, and S.~Shoaee, ``Extraordinarily long diffusion
  length in {PM6}:{Y6} organic solar cells,'' {\em Journal of Materials
  Chemistry A}, vol.~8, no.~16, p.~7854, 2020.
\newblock Publisher: Royal Society of Chemistry.

\bibitem{shannon_entropie_1948}
C.~E. Shannon, ``A mathematical theory of communication,'' {\em The Bell System
  Technical Journal}, vol.~27, no.~3, pp.~379--423, 1948.

\bibitem{blakesley_towards_2014}
J.~C. Blakesley, F.~A. Castro, W.~Kylberg, G.~F.~A. Dibb, C.~Arantes,
  R.~Valaski, M.~Cremona, J.~S. Kim, and J.-S. Kim, ``Towards reliable
  charge-mobility benchmark measurements for organic semiconductors,'' {\em
  Organic Electronics}, vol.~15, p.~1263, June 2014.

\bibitem{Cover_Knn_1967}
T.~Cover and P.~Hart, ``Nearest neighbor pattern classification,'' {\em IEEE
  Transactions on Information Theory}, vol.~13, no.~1, p.~21, 1967.

\bibitem{breiman_random_2001}
L.~Breiman, ``Random {Forests},'' {\em Machine Learning}, vol.~45, p.~5, Oct.
  2001.

\bibitem{Chen_XGboost_2016}
T.~Chen and C.~Guestrin, ``Xgboost: A scalable tree boosting system,'' in {\em
  Proceedings of the 22nd ACM SIGKDD International Conference on Knowledge
  Discovery and Data Mining}, KDD '16, (New York, NY, USA), p.~785, Association
  for Computing Machinery, 2016.

\bibitem{drucker_support_1996}
H.~Drucker, C.~J.~C. Burges, L.~Kaufman, A.~Smola, and V.~Vapnik, ``Support
  {Vector} {Regression} {Machines},'' in {\em Advances in {Neural}
  {Information} {Processing} {Systems}} (M.~C. Mozer, M.~Jordan, and
  T.~Petsche, eds.), vol.~9, MIT Press, 1996.

\bibitem{fujiwara2007spectroscopic}
H.~Fujiwara, {\em Spectroscopic ellipsometry: principles and applications}.
\newblock John Wiley \& Sons, 2007.

\bibitem{shuttle2008}
C.~G. Shuttle, B.~O’Regan, A.~M. Ballantyne, J.~Nelson, D.~D.~C. Bradley,
  J.~de~Mello, and J.~R. Durrant, ``{Experimental determination of the rate law
  for charge carrier decay in a polythiophene: Fullerene solar cell},'' {\em
  Applied Physics Letters}, vol.~92, p.~093311, 03 2008.

\bibitem{shuttle2010}
C.~G. Shuttle, R.~Hamilton, J.~Nelson, B.~C. O'Regan, and J.~R. Durrant,
  ``Measurement of charge-density dependence of carrier mobility in an organic
  semiconductor blend,'' {\em Advanced Functional Materials}, vol.~20, no.~5,
  pp.~698--702, 2010.

\end{thebibliography}
\bibliographystyle{ieeetr}

\clearpage
\onecolumngrid
\section{Supplementary material}
\textbf{Fabrication of PM6:DTY6 devices}

Materials: PM6 (95K) was purchased from Solarmer. DTY6 was provided by Prof. Lei Ying’s group at South China University of Technology (SCUT), China. SnO\textsubscript{2} nanoparticles (Product N-31) were received from Avataman. The o-Xylene solvent was purchassed from Sigma-Aldrich. All the materials were used as received without further purification

Device fabrication was based on the inverted devices with a configuration of ITO/SnO\textsubscript{2}/PM6:DT-Y6/MoO\textsubscript{3}/Ag.
First, the ITO substrates were cleaned in sequence in water, acetone, and Isopropanol, then dried with compress air. SnO\textsubscript{2} NPs were dispersed with ultrasonic treatment for 2 min and then filtered through 0.45 µm Polyamide (PA) filter before use. A 25 nm thickness of SnO\textsubscript{2} film was deposited on the ITO substrates by spin-coating; sequentially, the SnO\textsubscript{2} films were annealed at 200 °C for 30 min in air. Afterwards, active films with various D/A ratios (w/w), including 1:0, 0.85:0.15, 0.7:0.3, 0.55:0.45, 0.3:0.7, 0.15:0.85, and 0:1, were spun on the top of glass/ITO/SnO\textsubscript{2} in a nitrogen-filled glove box. For the ratios of 1:0, 0.85:0.15 and 0.7:0.3, the total concentration was 9 mg/ml in o-Xylene. For the ratios of 0.55:0.45, 0.3:0.7, 0.15:0.85, and 0:1, the total concentration was 18 mg/ml in o-Xylene. The thicknesses of all films were controlled by varying the spin speed. All films were annealed at 100°C/10 min in nitrogen atmosphere. Finally, all devices were completed by depositing 10 nm \textsubscript{3} and 100 nm Ag electrode through a mask with an opening area of 0.104 mm\textsuperscript{3} under $1\times 10^{-6}$ mbar.

\textbf{Optical measurement of PM6:DTY6}

For the optical constants, both refractive index $n$ and extinction coefficient $k$ are determined by spectroscopic ellipsometry (ME-L ellipsometer, Wuhan Eoptics Technology Co.). The samples were prepared on Si wafers under the same conditions used for device fabrication without additional post-processing. Spectroscopic ellipsometry measures $\Psi$ (related to the polarized light amplitude) and $\Delta$ (related to the polarized light phase) values, which are associated with the complex Fresnel reflection coefficients $r_\mathrm{s}$ (for s-wave) and $r_\mathrm{p}$ (for p-wave): 
\begin{equation*}
    \rho = \tan{\Psi}\exp{i\Delta} = \frac{r_\mathrm{p}}{r_\mathrm{s}}
\end{equation*}

After obtaining $\Psi$ and $\Delta$, we used the Cauchy model to fit $\Psi$ and $\Delta$ to determine the thicknesses of thin film samples on Si wafers, and further obtained optical constants of the materials through the fitting of Gaussian model and Tauc-Lorentz model \cite{fujiwara2007spectroscopic}. 

\textbf{Current--voltage characterization.}

A Keithley~236 SMU was used for voltage application and current measurement. AM1.5 illumination was provided by a Wavelabs LS-2 solar simulator. No aperture was used. The illumination was kept switched on for two seconds per measurement to prevent the sample temperature from increasing. We measured from reverse bias to forward bias with no fixed sweep speed due to enabled autoranging. Measurements were conducted in a nitrogen-filled glovebox.

\textbf{Intensity-modulated spectroscopy.}

Modulated and continuous illumination was provided by an Omicron A350 diode laser with a center wavelength of 515~nm. A Zurich Instruments MFLI lock-in amplifier with MF-IA, MF-MD, and MF-5FM options was used to measure sample current and voltage as well as providing voltage to modulate the laser. The illumination intensity was varied using neutral density filters mounted in a Thorlabs motorized filter wheel FW102C combined with a continuously variable neutral density filter wheel. For IMPS and IMVS measurements, the amplitude of modulated illumination was chosen to be 10\% of the bias illumination intensity to ensure small-signal excitation. Laser calibration was performed using a Newport 818-BB-21 biased silicon photodetector. 



\textbf{Transient Photovoltage}
Transient photovoltage measurements are collected on complete devices to characterise the charge carrier lifetime at different charge carrier densities in the device. To achieve that the device is kept under open circuit conditions with an LED bias light (ring of 6 cold and 6 warm white light LEDs) generating a background carrier density in the device. The LED light intensity is calibrated by using the J\textsubscript{sc} value obtained using the solar simulator. An additional laser pulse (532 nm, 5ns) Continuum Minilite Nd:YAG is used to provide a small voltage perturbation. Subsequently the voltage decays down to the steady state open circuit voltage. This voltage transient is recorded using the 1 M$\Omega$ input oscilloscope (Tektronix TDS3032B) and fitted with a single exponential. Finally the small perturbation lifetime is multiplied by the experimentally determined recombination order to yield the full charge carrier lifetime\cite{shuttle2008}.

\textbf{Charge-extraction}
For the charge extraction measurements, the same white light LEDs as used for the TPV measurements are used to illuminate the device. The device is kept under short circuit conditions and upon switching of the light, the decay of the current density from steady state short circuit to zero dark current is recorded by measuring the voltage drop across a 50$\Omega$ resistor connected to the 1M$\Omega$ input of an oscilloscope (Tektronix TDS3032B) and converting the voltage to a current transient using Ohm's Law. By integrating the current transient, the total carrier density can be calculated. This is used to calculate the effective mobility as previously described \cite{shuttle2010}.


\textbf{Details on training set generation}

The device is replicated in the drift diffusion simulation model OghmaNano.
20,000 copies with randomly generated device parameters are made. For each copy the JV-curves at the respective light intensities are simulated and saved together with the simulation results like charge carrier mobilities and recombination rates.
The range of simulation parameters used by the drift diffusion model is noted in Table \ref{tab:driftdiffusion}.
Each device is simulated at the respective intensities.
The Simulated JV-curves get sampled at (-2.0, -1.0, -0.1, 0.0, 0.1, 0.2, 0.3, 0.4, 0.5, 0.6, 0.7, 0.8, 0.9, 1.0, 1.4) V.

\begin{table}[H]
    \centering
    \begin{tabular}{llll}
    \hline
    Parameter & Min & Max & Units \\
    \hline
    $R_\mathrm{shunt}$ & $1.0$ & $10^{4}$ & $\unit{\Omega m^{2}}$\\
    $R_\mathrm{series}$ & $0.01$ & $100$ & $\unit{\Omega}$\\
    $\mu_\mathrm{free}^{e}$ & $10^{-8}$ & $10^{-4}$ & $\unit{m^{2}V^{-1}s^{-1}}$\\
    $\mu_\mathrm{free}^{h}$ & $10^{-8}$ & $10^{-4}$ & $\unit{m^{2}V^{-1}s^{-1}}$\\
    $N_\mathrm{trap}^{e}$ & $10^{22}$ & $10^{27}$ & $\unit{m^{3}}$ \\
    $N_\mathrm{trap}^{h}$ & $10^{22}$ & $10^{27}$ & $\unit{m^{3}}$ \\
    $E_{U}^{e}$ & $0.01$ & $0.3$ & $\unit{eV}$ \\
    $E_{U}^{h}$ & $0.01$ & $0.3$ & $\unit{eV}$ \\
    $\sigma_\mathrm{n,e}$ & $10^{-22}$ & $10^{-17}$ & $\unit{m^{-2}}$ \\
    $\sigma_\mathrm{p,h}$ & $10^{-22}$ & $10^{-17}$ & $\unit{m^{-2}}$ \\
    $\sigma_\mathrm{n,h}$ & $10^{-25}$ & $10^{-22}$ & $\unit{m^{-2}}$ \\
    $\sigma_\mathrm{p,e}$ & $10^{-25}$ & $10^{-22}$ & $\unit{m^{-2}}$ \\
    \hline
    \end{tabular}
    \caption{The range of simulation parameters used by the drift diffusion model: Resistance and mobility were varied for intensity dependent data set. For the prediction on the database, all above parameters were considered.}
    \label{tab:driftdiffusion}
\end{table}

\textbf{The Neural Network}

The Neural Network consists of an input layer, 4 dense layers with 200, 50 , 50 and 50 neurons respectively and an output layer.
The tensorflow hyperband optimisation algorithm was used to do the initial optimisation of the network topology. A full list of hyperparameters used can be found in Table \ref{tab:hyperparams}.

\begin{table}[H]
    \centering
    \begin{tabular}{ll}
    \hline
     Hyperparameter &  Value \\
     \hline
     Topology & 4 dense layers (200/50/50/50) \\
     Activation function & Sigmoid \\
     Optimisation function & ADAM-optimiser \\
     Learning rate & 0.01 \\
     Loss-function & Mean squared error (and R$^{2}$-value)\\
     Weight initialisation & Glorot-uniform initialisation\\
     training epoch & 500 with early stopping\\
     Batch size & 32\\
     \hline
    \end{tabular}
    \caption{Hyperparameters used for the ANN model for light intensity dependent prediction on JV-curves.}
    \label{tab:hyperparams}
\end{table}

\textbf{Feature and target normalisation}

For both feature and target normalisation we employed re scaling (min-max-normalisation) as follows
\begin{equation*}
    x_\mathrm{norm} = \frac{(x-x_\mathrm{min})}{(x_\mathrm{max}-x_\mathrm{min})}
\end{equation*}
The minimum and maximum values are always inferred from the whole training data set before splitting it into test and training set. For features, each light intensity is normalised on its own. The re-scaling projects all values in the simulated data set into the interval $[0.0,1.0]$. The minimum and maximum of the data set are stored and passed on to ensure the experimental data is re-scaled in a consistent way and the predicted values can be scaled back into their original value space.

\textbf{Training of the model}

For monitoring the training, we use the mean-squared-error as a loss function
\begin{equation*}
    MSE = \frac{1}{n}\sum^{n}_{i=0}\left(y_{i} - \hat{y}_{i}\right)^2
\end{equation*}
where $n$ is the number of outputs, $y$ is the predicted value of the ANN and $\hat{y}$ is the ground truth, also called label.
While the MSE enables to compare different models between each other, it is not very intuitive on its own. Therefore another metric is used in this work, the R2-score.
The R2-score is calculated by the fraction of residual sum of squares $SS_\mathrm{res}$ over the total sum of squares $SS_\mathrm{tot}$
\begin{equation*}
    R^2 = 1 - \frac{SS_\mathrm{res}}{SS_\mathrm{tot}} = \frac{\sum^{n}_{i=0}\left(y_{i} - \hat{y}_{i}\right)^2}{\sum^{n}_{i=0}\left(y_{i} - \bar{y}\right)^2}
\end{equation*}
with $\bar{y}$ being the mean of the predicted values. A perfect model will predict the observed value and therefore have $SS_\mathrm{res}=0$ and $R^2=1$. A model that always predicts $\bar{y}$ will have an R2-score of $R^2=0$ and a worse model will have a negative R2-score.
The best model (highest R2-score) is saved and used for the predictions on experimental data.

\textbf{Fabrication of evaporated devices}
Patterned indium tin oxide (ITO) glass substrates were pre-cleaned successively with detergent, acetone, de-ionized water and isopropyl alcohol and dried by nitrogen. The layers of the organics are thermally evaporated at ultra-high vacuum (base pressure < $10^{-7}$ mbar) on a glass substrate with a pre-structured ITO contact. All organic materials were purified 2-3 times by sublimation. The device area is defined by the geometrical overlap of the bottom and the top contact and equals 6 mm2. To avoid exposure to ambient conditions, the organic part of the device was covered by a small glass substrate, which is glued on top.

\textbf{Ease of parameter extraction}

\begin{figure}
    \centering
    \includegraphics[width = \textwidth]{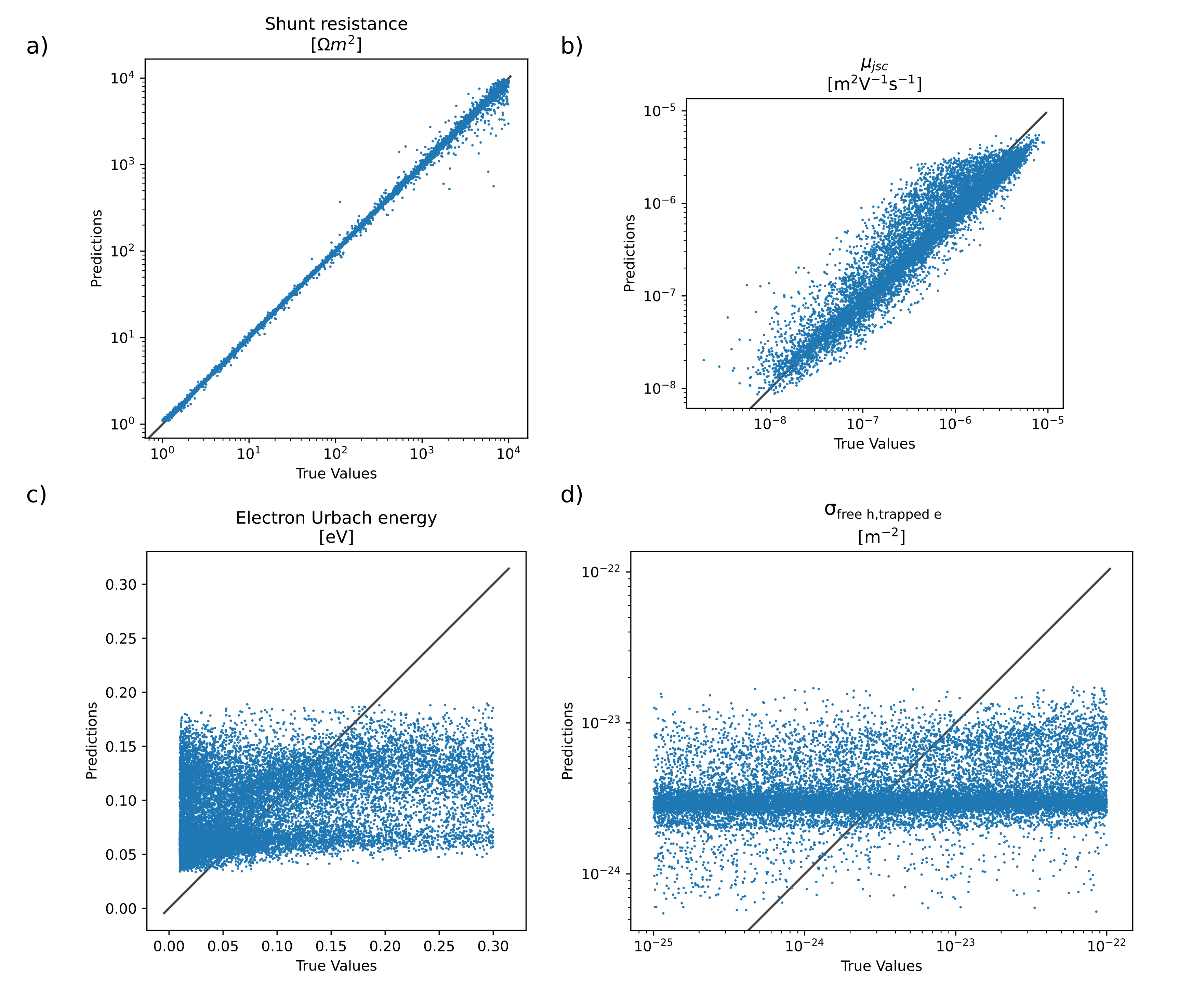}
    \caption{Confusion plots of the trained model predictions on the test-set for a) Shunt resistance, b) mobility at $j_\mathrm{sc}$ c) Urbach energy for electrons and d) SRH capture cross-section for free holes recombining with trapped electrons.}
    \label{fig:SI-parameters}
\end{figure}

One would think that in principle the machine learning model could be trained to predict all input parameters used in the generation of the dataset. However, some parameters are intrinsically harder to extract than others. Figure \ref{fig:SI-parameters} shows the confusion plots of the trained model predictions on the test-set. Each data point corresponds to one simulated solar cell device from the test-set of the training data. Plotted are the predictions of the machine learning model versus the true value (input value of the simulation). As one can see in Figure \ref{fig:SI-parameters} a the shunt resistance of the devices can be almost perfectly predicted from the JV-curves, therefore the data points lay on the diagonal where prediction equals true value. Similarly, other parameters such as series resistance, $V_\mathrm{oc}$, $j_\mathrm{sc}$, FF and PCE can be extracted very accurately by the model. It has to be noted, that these parameters could be extracted rather easily by a human as well. Other information that is harder to extract for humans such as charge carrier mobility (see Figure \ref{fig:SI-parameters} b) and lifetime can be accurately predicted by the model.

Mobility and recombination are in defined by other more microscopic parameters such as free charge carrier density $n_{free}$, traped charge carrier density $n_{trap}$, Urbach energies $E_\mathrm{U}$ and capture cross-sections $\sigma$ or mathematically put:

\begin{equation*}
    \mu = f(n_{free}, n_{trap}, E_\mathrm{U}, \sigma_\mathrm{n,e}, \sigma_\mathrm{p,h}, \sigma_\mathrm{n,h}, \sigma_\mathrm{p,e})
\end{equation*}

\begin{equation*}
    \tau = f(n_{free}, n_{trap}, E_\mathrm{U}, \sigma_\mathrm{n,e}, \sigma_\mathrm{p,h}, \sigma_\mathrm{n,h}, \sigma_\mathrm{p,e})
\end{equation*}

We found it easy to access $\mu$ and $\tau$ using the machine learning model but far harder to access the parameters upon which it depends. This includes trapping related parameters such as Urbach energy shown in Figure \ref{fig:SI-parameters} c or SRH capture cross-sections as shown in Figure \ref{fig:SI-parameters} d. 
This suggests that there is not enough information in the JV curve to independently extract  $n_\mathrm{trap}$, $E_\mathrm{U}$, $\sigma$. And one would need other experiments which contain more information to access this information (possibly temperature dependent measurements). Thus we define two types of parameters \emph{visible macroscopic} parameters which can be extracted using easily the ML and \emph{hidden microscopic} which although important and their influence can be measured they can not be directly measured themselves.

\end{document}